# The Growth of Supermassive Black Holes Across Cosmic Time

*Kirpal Nandra*

Astrophysics Group, Imperial College London, Blackett Laboratory, Prince Consort Road, London SW7 2AZ, UK. Tel: +44 207 594 5785, e-mail: k.nandra@imperial.ac.uk

J.A. Aird[1], D.M. Alexander[2], D.R. Ballantyne[3], X. Barcons[4], F.E. Bauer[5], T. Boller[6], W.N. Brandt[7], M. Brusa[6], A. Cattaneo[8], G. Chartas[7], A.L. Coil[9], A. Comastri[10], D.J. Croton[11], R. Della Ceca[12], M. Dickinson[13], A.C. Fabian[14], G.G. Fazio[15], F. Fiore[16], K.A. Flanagan[17], W.R. Forman[15], N. Gehrels[18], A. Georgakakis[19], I., Georgantopoulos[19], R. Gilli[10], G. Hasinger[20], P.F. Hopkins[21], A.E. Hornschemeier[18], R.J. Ivison[22], G. Kauffmann[23], A.R. King[24], A.M. Koekemoer[17], D.C. Koo[25], H. Kunieda[26], E.S. Laird[1], N.A. Levenson[27], Y. Li[15], P. Madau[25], T. Ohashi[28], K.A. Pounds[24], J.R. Primack[25], P. Ranall[10], G.R. Ricker[30], E.M. Rossi[31], O. Shemmer[32], R.S. Somerville[17], D. Stern[33], M. Stiavelli[17], H. Tananbaum[15], Y. Terashima[34], E. Treister[35], Y. Ueda[36], C. Vignali[10], M. Volonteri[37], M.G. Watson[26], N.E. White[18], S.D.M. White[23]

1 Astrophysics Group, Imperial College London, Blackett Laboratory, Prince Consort Road, London SW7 2AZ, UK
2 Department of Physics, Durham University, South Road, Durham, DH1 3LE, UK
3 Center for Relativistic Astrophysics, School of Physics, Georgia Institute of Technology, Atlanta, GA 30332, USA
4 Instituto de Física de Cantabria (CSIC-UC), Avenida de los Castros, 39005 Santander, Spain
5 Columbia Astrophysics Laboratory, Columbia University, 550 W. 120th St., Rm 1418, New York, NY 10027, USA
6 Max-Planck-Institut für extraterrestrische Physik, Giessenbachstrasse 1, D-85478 Garching, Germany
7 Department of Astronomy and Astrophysics, Penn State University, 525 Davey Lab, University Park, PA 16802, USA
8 Astrophysikalisches Institut Potsdam, an der Sternwarte 16, 14482 Potsdam, Germany
9 Department of Physics, University of California, San Diego, CA 92093, USA
10 INAF-Osservatorio Astronomico di Bologna, via Ranzani 1, I-40127 Bologna, Italy
11 Centre for Astrophysics & Supercomputing, Swinburne University, P.O. Box 218, Hawthorn, VIC3122, Australia
12 INAF - Osservatorio Astronomico di Brera, via Brera 28, 20121 Milan, Italy
13 National Optical Astronomy Observatory, 950 North Cherry Avenue, Tucson, AZ 85719, USA
14 Institute of Astronomy, Madingley Road, Cambridge CB3 0HA, UK
15 Harvard-Smithsonian Center for Astrophysics, 60 Garden Street, Cambridge, MA 02138, USA
16 INAF-Osservatorio Astronomico di Roma, via Frascati 33, 00040 Monte Porzio Catone (RM), Italy
17 Space Telescope Science Institute, 3700 San Martin Drive, Baltimore, MD 21218
18 NASA Goddard Space Flight Center, Greenbelt, MD 20771, USA
19 National Observatory of Athens, Institute of Astronomy,V. Paulou & I. Metaxa, Athens 15236, Greece
20 Max-Planck-Institute for Plasma Physics, Boltzmannstr. 2, D-85478 Garching, Germany
21 Department of Astronomy, University of California Berkeley, Berkeley, CA 94720, USA
22 UK Astronomy Technology Centre, Royal Observatory, Blackford Hill, Edinburgh EH9 3HJ, UK
23 Max-Planck Institut fuer Astrophysik, Karl-Schwarzschild Strasse 1, D-85748 Garching, Germany
24 University of Leicester, University Road, Leicester LE1 7RH, UK
25 UCO/Lick Observatory, University of California, Santa Cruz, CA 95064, USA
26 Department of Physics, Nagoya University, Furo-cho, Chikusaku, Nagoya 464-8602, Japan
27 Department of Physics and Astronomy, University of Kentucky, Lexington, KY 40506, USA
28 Department of Physics, Tokyo Metropolitan University, 1-1 Minami-Osawa, Hachioji, Tokyo 192-0397, Japan
29 RIKEN, Cosmic Radiation Laboratory, Hirosawa 2-1, Wakoshi, Saitama 351-0198, Japan
30 Kavli Institute for Astrophysics and Space Research, MIT, Cambridge, MA 02139-4307, USA
31 Racah institute for Physics, The Hebrew University, Jerusalem, 91904, Israel
32 Department of Physics, University of North Texas, Denton, TX 76203, USA
33 Jet Propulsion Laboratory, California Institute of Technology, Pasadena, CA 91109, USA
34 Department of Physics, Ehime University, Matsuyama, Ehime 790-8577, Japan
35 Institute for Astronomy, 2680 Woodlawn Drive, University of Hawaii, Honolulu, HI 96822, USA
36 Department of Astronomy, Kyoto University, Kyoto 606-8502, Japan
37 Department of Astronomy, University of Michigan, Ann Arbor, MI 48109, USA

# The Growth of Supermassive Black Holes Across Cosmic Time

*How did feedback from supermassive black holes shape the properties of galaxies?*

One of the main themes in extragalactic astronomy for the next decade will be the evolution of galaxies over cosmic time. Many future observatories, including JWST, ALMA, GMT, TMT and E-ELT will intensively observe *starlight* over a broad redshift range, out to the dawn of the modern Universe when the first galaxies formed. It has, however, become clear that the properties and evolution of galaxies are intimately linked to the growth of their central black holes. Understanding the formation of galaxies, and their subsequent evolution, will therefore be incomplete without similarly intensive observations of the *accretion light* from supermassive black holes (SMBH) in galactic nuclei. To make further progress, we need to chart the formation of typical SMBH at $z>6$, and their subsequent growth over cosmic time, which is most effectively achieved with X-ray observations. Recent technological developments in X-ray optics and instrumentation now bring this within our grasp, enabling capabilities fully matched to those expected from flagship observatories at longer wavelengths.

**The co-evolution of black holes and galaxies**

A major recent development in astrophysics has been the discovery of the relationship in the local Universe between the properties of galaxy bulges, and dormant black holes in their centers (Magorrian et al. 1998; Ferrarese & Merrit 2000; Gebhardt et al. 2000). These tight relationships represent definitive evidence for the co-evolution of galaxies and Active Galactic Nuclei (AGN). The remarkable implication of this is that some consequence of the accretion process on the scale of the black hole event horizon is able to influence the evolution of an entire galaxy. The main idea is that radiative and mechanical energy from the AGN regulates both star formation and accretion during periods of galaxy growth.

This kind of black-hole driven feedback is thought to be essential in shaping the first galaxies. Current models propose that mergers of small gas-rich proto-galaxies in deep potential wells at high redshift drive star formation and black hole growth (in proto-quasar active galaxies) until a luminous quasar forms. At this point, a black-hole driven wind evacuates gas from the nascent galaxy, limiting additional star formation and further black hole growth (Silk & Rees 1998; Fig. 1). Further episodes of merger-driven star formation, accretion, and feedback are expected to proceed through cosmic time. This provides a plausible origin for the M-$\sigma$ relation (e.g. King 2003), and explains many outstanding problems in galaxy evolution (e.g. Croton et al. 2005; Hopkins et al. 2006). Despite the intense current interest in this topic, and its great importance, direct evidence for widespread AGN feedback at high redshift is scarce and the details of the physical processes are unclear. It is thus certain to remain one of the most active topics in astrophysics during the next decade and beyond.

**The first supermassive black holes**

The very first stars formed in primordial structures where gravity was able to overpower the pressure of the ambient baryons, some hundred million years after the Big Bang. The first seed black holes (~100 $M_\odot$) are left behind as remnants of the most massive stars. The first galaxies, hosting these first black holes in their cores, are responsible for reionizing the Universe by z~10, as shown by WMAP. Still, the highest redshift galaxies and quasars currently known are all in the range *z*=6–7. To understand the inner workings of the first luminous sources we need to bridge the gap between the few known sources at this redshift, and the information we can extract from the microwave background.

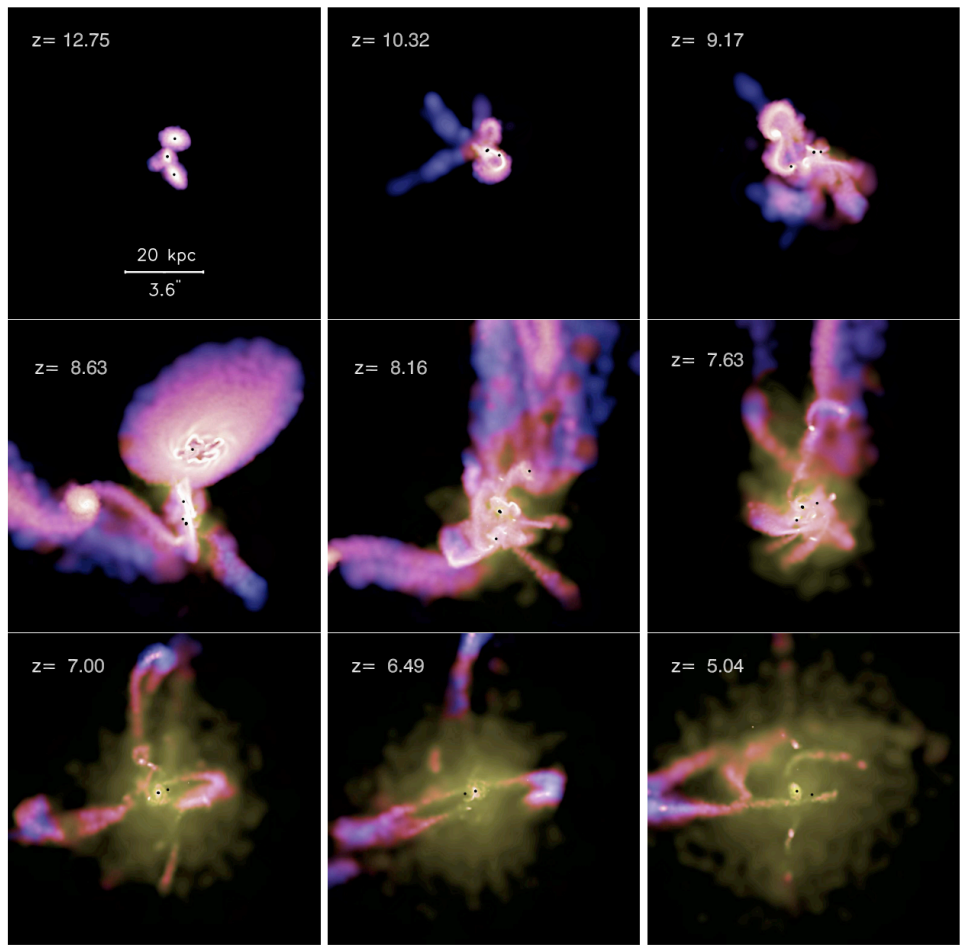

*Figure 1: Formation of a high-redshift quasar from hierarchical galaxy mergers as simulated by Li et al. (2007). Color shows gas temperature, and intensity shows gas density. Black dots represent black holes. Small, gas-rich galaxies merge in the deepest potential wells at high redshift, promoting star formation and black hole growth. At z ~ 7 to z ~ 5 a luminous quasar forms, associated with the most massive black hole. It drives a wind (yellow) that evacuates gas from the nascent galaxy.*

The known AGN population at *z*=6-7 consists of luminous optical quasars (e.g. Fan et al. 2003). Growing the extremely massive black holes required in <1 Gyr represents a challenge for theoretical models, because it requires Eddington-limited accretion over many folding times. Recent gas-dynamical cosmological simulations are nevertheless able to produce quasars with ~$10^9$ $M_\odot$ at $z$=6.5 through a rapid sequence of mergers in small groups of proto-galaxies (Li et al. 2007; Fig. 1). The growth is likely to proceed in a self-regulated manner owing to feedback with the progenitor host, with a period of intense star formation and obscured accretion preceding the optically bright quasar phase. The complex physics involved in such a scenario is, however, poorly understood.

It must also be borne in mind that these luminous QSOs, hosting among the most massive black holes (>$10^9$ $M_\odot$) in the Universe, are extremely rare. Typical AGN, which are of lower luminosity and often obscured, remain largely undiscovered. Uncovering such objects at $z$=6-7, and searching for them at even higher redshift holds the key to our understanding of this crucial phase in the development of the Universe. It is very likely that SMBH as massive as $10^6$ $M_\odot$ hosted by vigorously star forming galaxies, existed as early as $z$=10–11. X-ray observations offer a unique tool to discover and study the accretion light from moderate luminosity AGN at $z$=6-11, which are rendered invisible in other wavebands due to intergalactic absorption and dilution by their host galaxy.

**Obscured accretion and galaxy evolution**

The tight relation between galaxy bulges and black holes shows that star formation and accretion must have co-evolved throughout the history of the Universe. To understand this we need to uncover AGN over a broad range of redshifts, luminosities, and obscuration, to characterise the accretion history over the whole of cosmic time. In the theoretical framework discussed above, we expect most of the accretion at high redshift to be heavily obscured. Observational support for this comes from the deepest *Chandra* and *XMM-Newton* surveys, which are most likely missing a significant fraction of the total AGN population. At least 50% of the >6 keV background is still unresolved (e.g. Worsley et al. 2005). Population synthesis models (Comastri et al. 1995; Treister & Urry 2005; Gilli et al. 2007) predict the sources of the unresolved X-ray background to be heavily obscured, Compton-thick AGN ($N_H$>$10^{24}$ cm$^{-2}$). While there is a sizable population of these in the local Universe (Risaliti et al. 1999), their properties are basically unknown at larger distances, and their numbers are controversial (Treister et al. 2009). It has been suggested that Compton-thick AGN at z~2 may be hiding among infrared bright, optically faint galaxies (e.g. Daddi et al. 2007; Alexander et al. 2008; Fiore et al. 2008). Their average hard X-ray flux corresponds to an intrinsic X-ray luminosity >$10^{43}$ erg s$^{-1}$, so if there is a large population of such objects at cosmological redshifts, they could make a major contribution to the total accretion power (Fabian & Iwasawa 1999).

They could also radically alter our inferences regarding co-evolution scenarios. The leading hypothesis is that intense periods of star formation and black hole growth occur concurrently in the history of massive galaxies, possibly triggered by mergers (e.g. Sanders et al. 1988; Springel et al. 2005). Eventually, feedback from the AGN terminates star formation and, eventually, extinguishes the AGN itself. To test these models, and understand the process of co-evolution, we must determine the key properties of galaxies undergoing an active black hole growth phase, such as their morphologies, star formation rates, and star formation histories. Despite much recent progress, this has proved notoriously difficult. Even with the best current data, it is very challenging to disentangle emission from the galaxy from that associated with the AGN. Furthermore, given the considerations regarding Compton thick sources discussed above, it seems likely that at present we are far from a complete census of the AGN population.

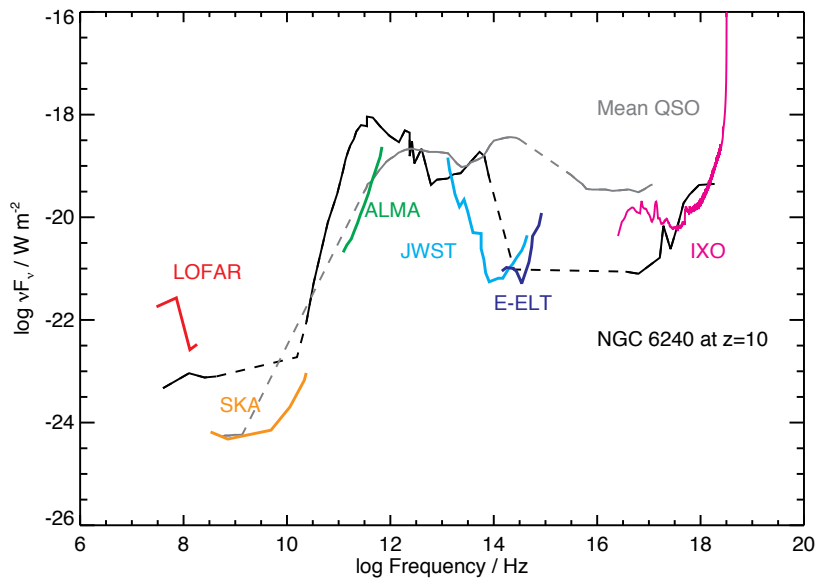

Figure 2: Sensitivity of future multiwavelength facilities to AGN at high $z$. An average QSO template, and the obscured starforming merger NGC 6240 are shown at z=10. Dashed sections indicate unobserved portions of the spectrum. Sensitivities assume 1 Ms 5$\sigma$ detections for IXO and equivalent 12h 1$\sigma$ detections for the other instruments. The emission of NGC 6240 is dominated by starlight in all but the X-ray band, so while the other facilities can examine the evolving galaxy, an X-ray observatory like IXO is required to reveal and characterise the AGN component.

AGN in the local Universe tend to be in massive bulges with evidence for an abundant available gas supply (Kauffmann et al. 2003). At $z$~1, the X-ray AGN population is found in largely quiescent, red, host galaxies also with undisturbed, bulge-dominated morphologies (e.g. Grogin et al. 2005; Nandra et al. 2007). There is thus little evidence either for intense, ongoing star formation or merging in the bulk of known AGN, contrary to the dominant theoretical framework. On the other hand, at higher redshifts ($z$~2) submm-selected galaxies show evidence for rapid co-eval, obscured star formation and black hole growth (Page et al. 2001; Alexander et al. 2005) and these FIR luminous galaxies are commonly mergers. These apparent contradictions can be understood if heavily obscured AGN represent a different phase of galaxy evolution. Indeed, models predict that the

Compton thick phase may be precisely at the stage where the AGN blows out gas from the galaxy, terminating star formation (Hopkins et al. 2006). Such objects are difficult to detect without extremely sensitive hard X-ray data. To understand AGN/galaxy co-evolution fully we must, therefore, complete the census of the AGN population by uncovering and characterising the properties of obscured objects at the peak of AGN and starforming activity at $z$=1-3.

**Direct observation of feedback at high redshift: the smoking gun?**

AGN feedback is seen directly in the local Universe in the cores of some clusters. The central AGN clearly influences the surrounding gas, preventing infall, cooling and star formation (McNamara et al. 2000; Fabian et al. 2003). This mode of feedback appears to solve the cooling flow problem (Peterson et al. 2001), and can explain the observed properties of massive cluster galaxies, which would be dramatically different if star formation were to proceed unabated (Croton et al. 2005). Outside these massive haloes and at high redshift, where most galaxy building occurs, alternative modes of AGN feedback involving powerful winds have been proposed (e.g. Hopkins et al. 2006). To understand these we need to design experiments to observe feedback directly at $z$=1-3, at the peak of starforming and AGN activity in the Universe.

X-ray observations are uniquely powerful at revealing the conditions in the immediate vicinity of SMBH, including winds. When a black-hole driven wind lies along the line of sight, it causes ionized and/or neutral X-ray absorption via metal edges and lines; perhaps surprisingly, metals are known to be abundant in AGN out to the highest redshifts. Critically, the hotter X-ray absorbing component of the wind likely carries more mass and energy than components detected at longer wavelengths such as the rest-frame UV. X-ray measurements are thus essential for assessing the level of feedback. *XMM-Newton* and *Chandra* have shown that ionized outflows are common in nearby AGN. These are typically not powerful enough to influence the galaxy scale, but there are cases where much larger velocities and mass flows are implied (e.g. Pounds et al. 2003). At higher redshifts, AGN are usually too faint for sensitive X-ray spectroscopy using current instrumentation. Using the magnifying power of gravitational lensing, however, evidence has been presented for massive, high velocity outflows in a few rare cases at high $z$ (Chartas et al. 2002). The requirement now is to observe black hole winds in typical AGN at $z=1–3$, where the majority of galaxy growth occurs.

**Tackling the problem in the next decade**

Exploring galaxy evolution is an inherently multiwavelength problem. The premier current surveys (e.g. AEGIS, COSMOS, GOODS) use large investments of time from essentially all of the most sensitive space and ground-based facilities (e.g. *Chandra/XMM-Newton*, HST, Keck, SCUBA, Spitzer, VLA). Future experiments from the radio-through-optical (e.g. SKA, ALMA, JWST, E-ELT/GMT/TMT), will reach the highest redshifts, probing the evolution of the stellar component of galaxies out to the epoch when the first objects were born. Determining the role of black hole accretion in this process requires X-ray observations of comparable sensitivity (Fig. 2). New technological developments in X-ray optics and instrumentation now make feasible the required performance, and this provides one of the main science drivers for the *International X-ray Observatory* (IXO).

*The first supermassive black holes:* the highest redshift QSOs known have been discovered in wide-field optical surveys (e.g. SDSS; Fan et al. 2001). These are fascinating sources, but it is crucial to bear in mind that they are among the most extreme and unusual objects in the Universe. Large area optical surveys will be continued with, e.g. PAN-STARRS, LSST, VISTA, and perhaps JDEM/EUCLID, which will discover many more high z QSOs. The optical surveys are, however, fundamentally limited to objects in which the AGN outshines the galaxy. For this reason, X-ray observations can probe much lower bolometric luminosities than the optical. Current deep X-ray surveys probe factors of 100-1000 fainter down the luminosity function than SDSS, at or around L* at $z$=1-3, where typical objects reside and the bulk of the accretion power is produced. We furthermore expect the majority of AGN at high redshift to be heavily obscured by gas and dust, where the accretion light is rendered invisible in the optical but detectable at X-ray energies. X-ray observations are thus absolutely essential in both the discovery and characterization of *typical* accreting black holes at high redshift.

To detect typical AGN at z≥7 and investigate their growth requires an unprecedented combination of large throughput, good angular resolution and large field-of-view in the X-ray regime. Current surveys are unable to probe to sufficient depth over a wide enough area to find such objects in significant number, but this problem will be solved with IXO. Followup of faint IXO sources with JWST and/or ALMA will provide the necessary identification of the highest *z* AGN. Note that even though we expect ALMA and/or JWST to be able to detect the host galaxies of these objects, sensitive X-ray observations are needed to disentangle the power associated with black hole accretion from that due to star formation. Reasonable extrapolations of the best estimates of the X-ray luminosity function at high redshift (Aird et al. 2008; Brusa et al. 2009) predict a handful of z~7 AGN in the current deepest *Chandra* surveys (CDF-N, CDF-S and AEGIS). This is consistent with observations of optically invisible X-ray sources (EXOs; Koekemoer et al. 2004; note that JWST/ALMA may confirm these as being at $z$>7). IXO can reach the CDF depth in <1/10$^{th}$ of the *Chandra* exposure time, with a slightly larger FoV, more uniform sensitivity across the field, and vastly greater X-ray photon statistics. It will thus revolutionise this field, likely yielding many 10s to 100s of moderate luminosity AGN at z≥7 and pushing to $z$=8-10. Semi-analytic models predict numbers of ultra-high *z* AGN that differ by several orders of magnitude (cf. Rhook & Haehnelt 2008; Marulli et al. 2008), based on differing evolutionary scenarios. IXO will discriminate decisively between them.

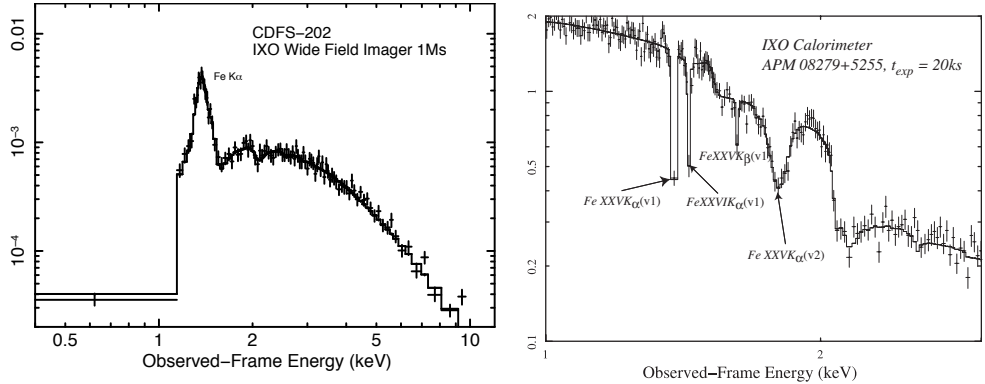

*Figure 3: Simulated spectra of (left) the z=3.7 Compton thick AGN CDFS-202 (Norman et al. 2002), in a 1Ms exposure with the IXO WFI; Parameters are based on the deep XMM observation of the CDF-S (PI: Comastri) (right) the characteristic X-ray BAL signatures of the fast (~0.2c) outflow in APM 08279+5255 (Chartas et al. 2002), which may be in the "blowout phase", as seen by the IXO/XMS calorimeter in just 20ks.*

*Obscured accretion:* X-ray selection has provided the most robust AGN samples to date, but finding the most obscured objects has proved difficult. Mid-IR selection is promising, but has not yet yielded samples that are both reliable and complete. Future hard X-ray (>10 keV) imaging (*NUSTAR, ASTRO-H, Simbol-X*) will provide a step forward in revealing AGN with column densities of ~few × $10^{24}$ $cm^{-2}$. Interestingly, hard X-ray surveys have revealed surprisingly few Compton thick AGN in the local universe (Tueller et al. 2008; Treister et al. 2009). They may be more rare than previously thought, or more heavily obscured. At the highest column densities, even the 10-40 keV light is suppressed (by a factor ~10 at $N_H=10^{25}$ $cm^{-2}$), leaving the AGN visible only in scattered X-rays. The spectral sensitivity of IXO in the 2-10 keV band will reveal the telltale intense iron K$\alpha$ emission characteristic of a Compton reflection dominated source (Fig. 3), and can be combined simultaneously with hard X-ray data (of unprecedented sensitivity if the goal angular resolution is achieved).

*Direct observations of feedback:* The same combination of unprecedented throughput and spectroscopic capability will open up the distant X-ray Universe to spectroscopy for the first time. Inter alia, this will enable the first significant samples of autonomously determined X-ray redshifts, and a detailed examination of any changes in the accretion process throughout cosmic time. For the present discussion, we highlight the extraordinary sensitivity of IXO to absorption features due to outflows, via iron K$\alpha$ and many other metal species. Detection of powerful winds in AGN hosts at the major epoch of galaxy formation is the "smoking gun" of feedback, and detailed spectroscopy will reveal the velocity, column density, metallicity and ionization structure of the outflows (Fig. 3) for detailed comparison with physical models, which are at present completely unconstrained.

Reference list and further information can be found at:
http://astro.imperial.ac.uk/research/ixo/smbh_growth.shtml